\documentclass[twoside,a4paper,11pt]{sea10}
\usepackage{graphicx}
\usepackage{hyperref}
\usepackage{movie15}
\begin{document}
\pagenumbering{arabic}
\pagestyle{myheadings}
\thispagestyle{empty}
{\flushleft\includegraphics[width=\textwidth,bb=58 650 590 680]{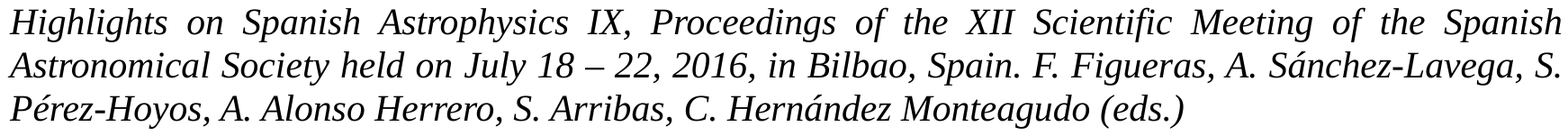}}
\vspace*{0.2cm}
\begin{flushleft}
{\bf {\LARGE
%
Neutron-capture element abundances in the planetary nebula NGC 5315 from deep high-resolution optical and near-IR spectrophotometry 
%
}\\
\vspace*{1cm}
%
S. Madonna,$^{1,2}$
J. Garc\'{i}a-Rojas,$^{1,2}$
N. C. Sterling$^{3}$
and V. Luridiana$^{1,2}$
%
}\\
\vspace*{0.5cm}
%
$^{1}$Instituto de Astrof{\'i}sica de Canarias, E-38205 La Laguna, Tenerife, Spain\\
$^{2}$Universidad de La Laguna, Dpto. Astrof{\'i}sica, E-38206 La Laguna, Tenerife, Spain\\
$^{3}$Department of Physics, University of West Georgia, 1601 Maple Street, Carrollton, GA 30118, USA
%
\end{flushleft}
%
\markboth{
Neutron-capture elements in NGC 5315
}{ 
%
Madonna et al. 2016
%
}
\thispagestyle{empty}
\vspace*{0.4cm}
\begin{minipage}[l]{0.09\textwidth}
\ 
\end{minipage}
\begin{minipage}[r]{0.9\textwidth}
\vspace{1cm}

\newcommand{\elecd}{$n_{\rm e}$}
\newcommand{\te}{$T_{\rm e}$}
\newcommand{\hb}{H$\beta$}
\newcommand{\ha}{H$\alpha$}
\newcommand{\fci}{[C~{\sc i}]}
\newcommand{\iii}{{\sc iii}}
\newcommand{\fniii}{[N~{\sc iii}]}
\newcommand{\foi}{[O~{\sc i}]}
\newcommand{\foii}{[O~{\sc ii}]}
\newcommand{\foiii}{[O~{\sc iii}]}
\newcommand{\fsi}{[S~{\sc i}]}
\newcommand{\fsii}{[S~{\sc ii}]}
\newcommand{\fsiii}{[S~{\sc iii}]}
\newcommand{\fnitroi}{[N~{\sc i}]}
\newcommand{\fnii}{[N~{\sc ii}]}
\newcommand{\sfmgi}{Mg~{\sc i}]}
\newcommand{\mgi}{Mg~{\sc i}}
\newcommand{\mgii}{Mg~{\sc ii}}
\newcommand{\fariii}{[Ar~{\sc iii}]}
\newcommand{\fariv}{[Ar~{\sc iv}]}
\newcommand{\farv}{[Ar~{\sc v}]}
\newcommand{\fcaii}{[Ca~{\sc ii}]}
\newcommand{\fclii}{[Cl~{\sc ii}]}
\newcommand{\fcliii}{[Cl~{\sc iii}]}
\newcommand{\fcliv}{[Cl~{\sc iv}]}
\newcommand{\fcrii}{[Cr~{\sc ii}]}
\newcommand{\fcriii}{[Cr~{\sc iii}]}
\newcommand{\fcriv}{[Cr~{\sc iv}]}
\newcommand{\fneiii}{[Ne~{\sc iii}]}
\newcommand{\fneiv}{[Ne~{\sc iv}]}
\newcommand{\fnev}{[Ne~{\sc v}]}
\newcommand{\fkriii}{[Kr~{\sc iii}]}
\newcommand{\fkriv}{[Kr~{\sc iv}]}
\newcommand{\fkrv}{[Kr~{\sc v}]}
\newcommand{\fkiv}{[K~{\sc iv}]}
\newcommand{\fkv}{[K~{\sc v}]}
\newcommand{\fkvi}{[K~{\sc vi}]}
\newcommand{\fxeiii}{[Xe~{\sc iii}]}
\newcommand{\fxeiv}{[Xe~{\sc iv}]}
\newcommand{\fxevi}{[Xe~{\sc vi}]}
\newcommand{\frbiv}{[Rb~{\sc iv}]}
\newcommand{\frbv}{[Rb~{\sc v}]}
\newcommand{\fnaiv}{[Na~{\sc iv}]}
\newcommand{\fniqi}{[Ni~{\sc i}]}
\newcommand{\fniqii}{[Ni~{\sc ii}]}
\newcommand{\fniqiii}{[Ni~{\sc iii}]}
\newcommand{\fniqiv}{[Ni~{\sc iv}]}
\newcommand{\fniqv}{[Ni~{\sc v}]}
\newcommand{\ffiv}{[F~{\sc iv}]}
\newcommand{\fcuv}{[Cu~{\sc v}]}
\newcommand{\fcuvi}{[Cu~{\sc vi}]}
\newcommand{\fcoiv}{[Co~{\sc iv}]}
\newcommand{\fcovi}{[Co~{\sc vi}]}
\newcommand{\fcav}{[Ca~{\sc v}]}
\newcommand{\fmniii}{[Mn~{\sc iii}]}
\newcommand{\fmniv}{[Mn~{\sc iv}]}
\newcommand{\fmnv}{[Mn~{\sc v}]}
\newcommand{\fmnvi}{[Mn~{\sc vi}]}
\newcommand{\sffeii}{Fe~{\sc ii}]}
\newcommand{\ffeii}{[Fe~{\sc ii}]}
\newcommand{\ffeiii}{[Fe~{\sc iii}]}
\newcommand{\ffeiv}{[Fe~{\sc iv}]}
\newcommand{\ffev}{[Fe~{\sc v}]}
\newcommand{\ffevi}{[Fe~{\sc vi}]}
\newcommand{\ffevii}{[Fe~{\sc vii}]}
\newcommand{\fpii}{[P~{\sc ii}]}
\newcommand{\piv}{P~{\sc iv}}
\newcommand{\fseii}{[Se~{\sc ii}]}
\newcommand{\fseiii}{[Se~{\sc iii}]}
\newcommand{\fseiv}{[Se~{\sc iv}]}
\newcommand{\fbriii}{[Br~{\sc iii}]}
\newcommand{\oiii}{O~{\sc iii}}
\newcommand{\oiv}{O~{\sc iv}}
\newcommand{\ov}{O~{\sc v}}
\newcommand{\nitroi}{N~{\sc i}}
\newcommand{\nii}{N~{\sc ii}}
\newcommand{\niii}{N~{\sc iii}}
\newcommand{\niv}{N~{\sc iv}}
\newcommand{\nv}{N~{\sc v}}
\newcommand{\sili}{Si~{\sc i}}
\newcommand{\silii}{Si~{\sc ii}}
\newcommand{\siliii}{Si~{\sc iii}}
\newcommand{\siliv}{Si~{\sc iv}}
\newcommand{\oi}{O~{\sc i}}
\newcommand{\oii}{O~{\sc ii}}
\newcommand{\ci}{C~{\sc i}}
\newcommand{\cii}{C~{\sc ii}}
\newcommand{\ciii}{C~{\sc iii}}
\newcommand{\civ}{C~{\sc iv}}
\newcommand{\cv}{C~{\sc v}}
\newcommand{\cvi}{C~{\sc vi}}
\newcommand{\nei}{Ne~{\sc i}}
\newcommand{\neii}{Ne~{\sc ii}}
\newcommand{\neiii}{Ne~{\sc iii}}
\newcommand{\neiv}{Ne~{\sc iv}}
\newcommand{\nev}{Ne~{\sc v}}
\newcommand{\sii}{S~{\sc ii}}
\newcommand{\siii}{S~{\sc iii}}
\newcommand{\niqi}{Ni~{\sc i}}
\newcommand{\niqii}{Ni~{\sc ii}}
\newcommand{\fciii}{C~{\sc iii}]}
\newcommand{\cli}{Cl~{\sc i}}
\newcommand{\cliii}{Cl~{\sc iii}}
\newcommand{\fei}{Fe~{\sc i}}
\newcommand{\feii}{Fe~{\sc ii}}
\newcommand{\feiii}{Fe~{\sc iii}}
\newcommand{\feiv}{Fe~{\sc iv}}
\newcommand{\ari}{Ar~{\sc i}}
\newcommand{\arii}{Ar~{\sc ii}}
\newcommand{\ariii}{Ar~{\sc iii}}
\newcommand{\ariv}{Ar~{\sc iv}}
\newcommand{\ki}{K~{\sc i}}
\newcommand{\hi}{H\,{\sc i}}
\newcommand{\hii}{H~{\sc ii}}
\newcommand{\di}{D\,{\sc i}}
\newcommand{\hei}{He~{\sc i}}
\newcommand{\heii}{He~{\sc ii}}
\newcommand{\mc}{\multicolumn}
\newcommand\mcnd{\multicolumn{1}{c}{\nodata}}

\section*{Abstract}{\small
%
We have done a spectroscopical analysis of the type I planetary nebula (PN) NGC\,5315, through high-resolution (R$\sim$40000) optical spectroscopy with UVES at the 8.2m Very Large Telescope, and medium-resolution (R$\sim$4800) near-IR spectroscopy with FIRE at the 6.5m Magellan Baade telescope, covering a wide spectral range from 0.31 $\mu$m to 2.50$\mu$m. The main aim of this work is to investigate the slow neutron(n)-capture process (the s-process) in the Asymptotic Giant Branch (AGB) star progenitor of a type I PNe. We detected and identified about 700 features, including lines from the n-capture elements Kr, Se, and possibly Br and Xe. We compute physical conditions using line ratios of common ions. Ionic abundances are computed for the species with available atomic data. We calculate total abundances using recent ionization correction factors (ICFs) or by summing ionic abundances. Our results for common elements are in good agreement with previous works on the same object.
We do not find a substantial \emph{s}-process enrichment in NGC\,5315, which is typical for type I PNe.   
\normalsize}
\end{minipage}
%
%
%
\section{Introduction \label{intro}}

Neutron(\emph{n})-capture elements (Z$>$30) are formed in the \emph{s}-process, where iron-peak nuclei experience slow neutron captures alternated with $\beta$ decays to form heavier elements. The \emph{s}-process is activated in AGB stars (1--8 M$_{\odot}$), in the intershell region between the H- and He-burning shells. The neutron flux is generated by $\alpha$-captures onto $^{13}$C (or $^{22}$Ne in AGB stars with mass $>3-4$ M$_{\odot}$). As $^{13}$C produces a higher time integrated neutron flux than $^{22}$Ne, the stellar mass plays a role in determining the final \emph{n}-capture element abundance distribution. The study of the nebular abundance of \emph{n}-capture elements is crucial to understand the physical conditions in stellar interiors and the nucleosynthetic histories of stellar populations (e. g., \cite{karakaslattanzio14}).

Neutron-capture elements were not identified in any astrophysical nebula until 1994, when \cite{pequignotbaluteau94} identified emission lines of Br, Kr, Rb, Xe, Ba, and possibly other heavy species in the bright PN NGC 7027. Since then, many detenctions of \emph{n}-capture element emission lines have been reported by several authors in the near-infrared (\hbox{e. g.}, \cite{sterlingdinerstein08}; \cite{sterlingetal16}), UV (\cite{sterlingetal02}), and optical (\hbox{e. g.}, \cite{sharpeeetal07, garciarojasetal12, garciarojasetal15}) spectra of PNe. The number of \emph{n}-capture element detections have spurred atomic data determinations that allow for new and accurate ionization correction factors (ICFs).  Deep, high-resolution spectra of PNe with large telescopes such as the VLT enable the detection of multiple ions of \emph{n}-capture elements, which further improve the accuracy of abundance determinations (\cite{garciarojasetal15}).

In this work we study NGC 5315, a type I PN. The high concentration of N and He, and the multipolar morphology of this kind of objects suggest a massive progenitor star \hbox{(M $>3-4$ M$_{\odot}$)}. Furthermore, there are statistical evidences that Type I and bipolar PNe come from a younger and more massive stellar population. However, three warning should be made: i) the total abundance of N from optical spectra is uncertain due to the large and uncertain ICFs (\cite{delgadoingladaetal15}); ii) extra mixing events during during the RGB and AGB phase may produce N and He, enhancing the final abundances in progenitors with masses lower than 3 M$_{\odot}$ (\cite{nollettetal03}), and iii) the peculiar morphology may be due to the presence of a close binary central star (\cite{balickfrank02}), where the interaction with the companion may truncate the AGB lifetime hampering the \emph{s}-process. Taking all this considerations into account even progenitor stars with masses $<$ 3--4 M$_{\odot}$ may generate type I PNe. The \emph{s}-process in the progenitors of type I PNe has been studied just for a few objects (\cite{sterlingdinerstein08}), in which a clear evidence of a little or absent \emph{s}-process enrichment is found. However, more observational constraints for type I PNe are needed to improve the theoretical models in the understanding of the physics of the stellar interiors and nucleosynthesis in their progenitor stars (\cite{karakaslattanzio14}).

\section{Data and Analysis}
\begin{figure}
\includegraphics[scale=0.45]{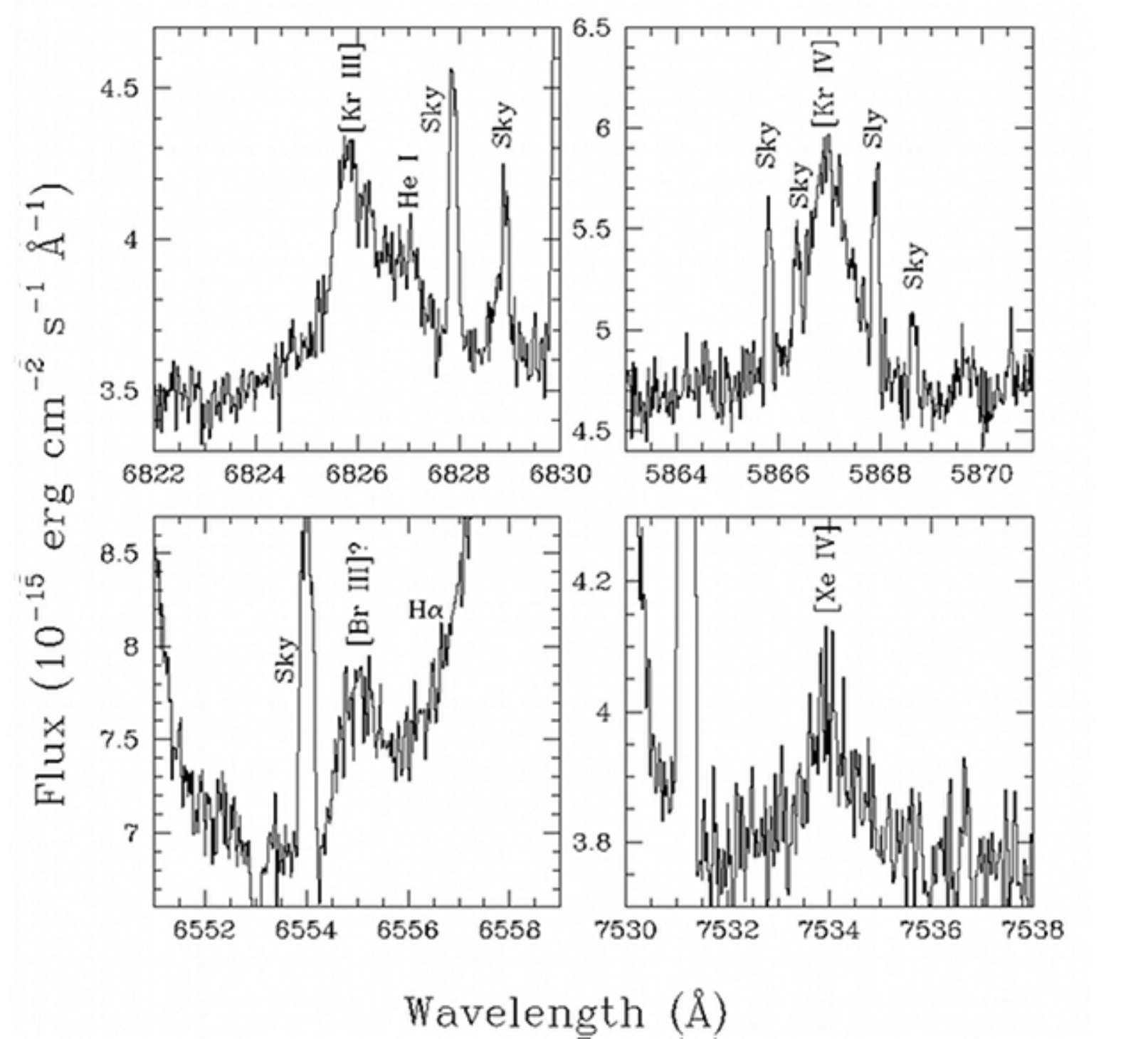} 
\includegraphics[scale=0.44]{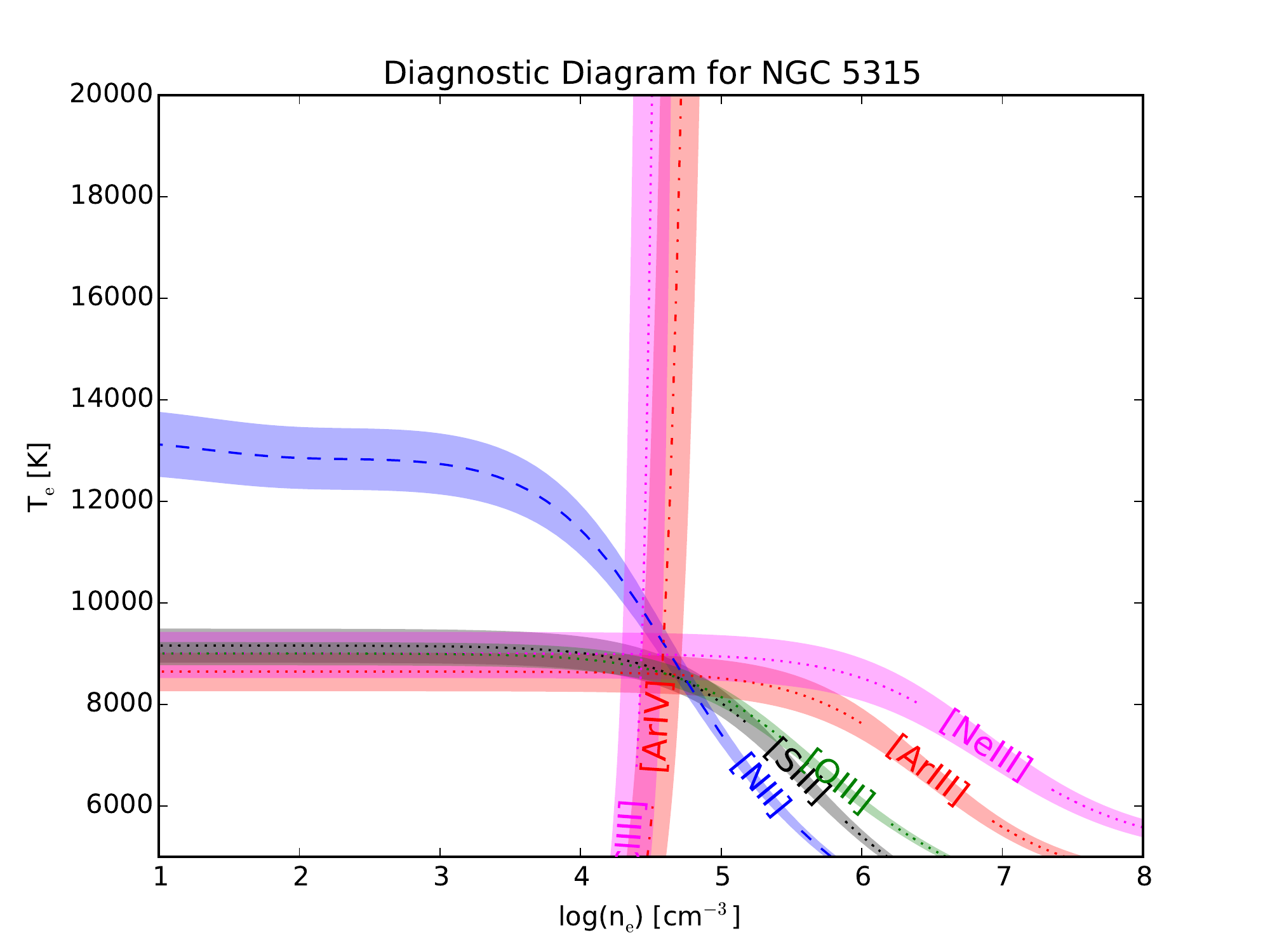} 
\caption{\label{fig1} Left panel: Portion of the spectra showing lines profiles of neutron-capture ion features. Right panel: Diagnostic diagram for the complete set of ionization zones for NGC 5315.
}
\end{figure}

In this work, we analyze the PN NGC 5315 joining the high-resolution (R$\sim$40000) optical spectra observed with UVES, attached to the 8.2m Kueyen (UT2) VLT at Cerro Paranal Observatory (Chile), with the medium-resolution (R$\sim$4800) NIR spectra taken with the Folded-port InfraRed Echellette (FIRE) spectrograph attached to the 6.5m Magellan Baade Telescope (MBT) located at Las Campanas Observatory in Chile. We detect, identify and measure about 700 lines belonging to several ions of different species. Some \emph{n}-capture ion line identifications are shown in figure \ref{fig1}. Lines were measured using the splot routine in IRAF, integrating all the flux of a given line  between two given limits and over a local continuum estimated by eye. In some cases of very tight blends including lines of interest we use the predicted intensities from a detailed Cloudy (\cite{ferlandetal13}) model. Physical conditions and ionic chemical abundances were computed using PyNeb (\cite{luridianaetal15}). We computed physical conditions (T$_{e}$ and n$_{e}$) using diagnostic line ratios of common ions (O$^{+}$, O$^{2+}$, S$^{+}$, S$^{2+}$, Cl$^{2+}$, Ar$^{2+}$, Ar$^{3+}$). The selected T$_{e}$ and n$_{e}$ diagnostic line ratios are shown in figure \ref{fig1}. Ionic abundances were calculated for the ions with available atomic data. Finally, to compute total abundances we used recent ionization correction factors (ICFs) developed by \cite{delgadoingladaetal14} for common elements, and by \cite{sterlingetal15} for the \emph{n}-capture elements Kr and Se. 
 
\begin{table}[ht] 
\caption{Kr and Se total abundances} 
\center
\begin{minipage}{0.8\textwidth}
\center
\begin{tabular}{ccc} 
\hline\hline 
ICF$^{a}$  &(12 + log(Kr/H)) & (12 + log(Se/H))  \\   
\hline 
eq. 1  & 3.61$\pm$0.10 & --\\
  eq. 2 &   3.84$\pm$0.25 & -- \\                      
  eq. 3 &4.16$\pm$0.14 & --\\
   eq. 4 & 3.60$\pm$0.10 & -- \\
   eq. 7 & -- & 3.88$\pm$0.26 \\
     eq. 8 & -- & 2.67$\pm$0.14 \\
       eq. 9 & -- & 3.48$\pm$0.16 \\
       \hline
       \multicolumn{3}{l}{$^a$ ICFs provided by \cite{sterlingetal15}}\\
\end{tabular} 
\end{minipage}
\label{s_abu} 
\end{table}

\section{Discussion and conclusions}

\cite{sterlingetal15} computed detailed ICFs for the \emph{n}-capture elements Kr and Se, through a large grid of Cloudy models spanning a wide range of physical conditions such as temperature and luminosity of the central star, and density and chemical composition of the PN. They propose 3 ICFs for Se and 6 ICFs for Kr. Thanks to the wide spectral range covered by our observations we have tested for the first time the complete set of Se ICFs. We considered as representative of Se abundance the result found with the ICF proposed in their Equation 9, in which two different ions (Se$^{2+}$ and Se$^{3+}$) are involved. The derived elemental Se abundances from Equations 7--9 of \cite{sterlingetal15} disagree, though the abundance from Equation 7 is uncertain due to the relatively high uncertainties of the Ar and Se$^{2+}$ abundances, and Equation 8 is the most uncertain among the ionization correction schemes used for Se (\cite{sterlingetal15}). Observations of [Se III] 1.0995 $\mu$m in additional PNe are needed to more rigorously test the Se ICFs of \cite{sterlingetal15}. For Kr, we could test only 4 ICFs, and as representative abundance we took an average of the abundances found with the ICFs of eq. 1 and 4, because the ICF of their eq. 2 show high uncertainties, due to the relatively large Ar abundance uncertainties, and the ICF of their eq. 3 is not suitable for low ionization PNe as NGC\,5315. The results are shown in table \ref{s_abu}.

\begin{table}[ht] 
\caption{Neutron-Capture Element Abundances} 
\center
\begin{minipage}{0.8\textwidth}
\center
\begin{tabular}{cr} 
\hline\hline 
&NGC\,5315 \\   
\hline 
		 $\lbrack$Se/Ar$\rbrack$& -0.20$\pm$0.26  \\
		 $\lbrack$Kr/Ar$\rbrack$& 0.05$\pm$0.26 \\
		 $\lbrack$Br/Ar$\rbrack$& 0.14:/0.68:$^{a}$  \\
		  $\lbrack$Rb/Ar$\rbrack$& $<$0.03  \\
		  $\lbrack$Xe/Ar$\rbrack$& 0.83:  \\
       \hline
        \multicolumn{2}{l}{$^a$ Considering the line at 6131 $\AA$/line at 6556 $\AA$}\\
\end{tabular} 
\end{minipage}
\label{s_enrich} 
\end{table}

In this work we do not find evidences of \emph{s}-process enrichment in NGC 5315. In table \ref{s_enrich}, we show the logarithmic difference between nebular and solar ratio abundances, using as reference element Ar, which is not affected by nucleosynthesis during the life of the stars. As proposed by \cite{sterlingdinerstein08} a threshold value of 0.2--0.3 dex establishes if the progenitor star has undergone a substantial {\emph s}-process enrichment. The most accurate abundance determinations are for Kr and Se, for which advanced ICFs are available. As is clearly seen in table \ref{s_enrich}, no enrichment is found for such elements. We neither found enrichment for Rb and Br, although the lines detected for such species are very uncertain (or an upper limit in the case of Rb) and the results must be taken with caution. The upper limit found for the enrichment of Rb is [Rb/Ar]$<$0.03, while in the case of Br we report the calculations made with two different lines of Br$^{2+}$. The fainter line at 6131 $\AA$ gives no enrichment, while the brighter line at 6556 $\AA$ suggests an unexpected enrichment. The detection of each line is marginal, and the brighter line at 6556 $\AA$ may be blended with an unknown feature. 

These results lead to a set of possible conclusions. i) Type I PNe may be generated by massive progenitor stars (\hbox{M $>3-4$ M$_{\odot}$}) and the absence of Rb enrichment suggests the upper limit \hbox{M $<$ 6 M$_{\odot}$} (\cite{karakaslugaro16}). This result agrees with \cite{sterlingetal15}, who found no significant \emph{s}-process enhancements in Type I PNe, possibly due to dilution of enriched material into the massive envelopes of the progenitor AGB stars.
ii) Binary interactions could have prematurely ended the AGB phase of the progenitor before third dredge-up could occur, resulting in no \emph{s}-process or C enrichment.  This conclusion is supported by the central star radial velocity variations found by \cite{manicketal15}, which may be due to a binary companion.
iii) Alternatively, our abundance results are consistent with a single, approximately solar-metallicity progenitor with mass $M < 1.5 \mathrm{M}_{\odot}$.  Such stars are not sufficiently massive to experience third dredge-up, and the He and N enrichments could be due to extra mixing processes \cite{nollettetal03}.

\begin{figure}
\center
\includegraphics[scale=0.5]{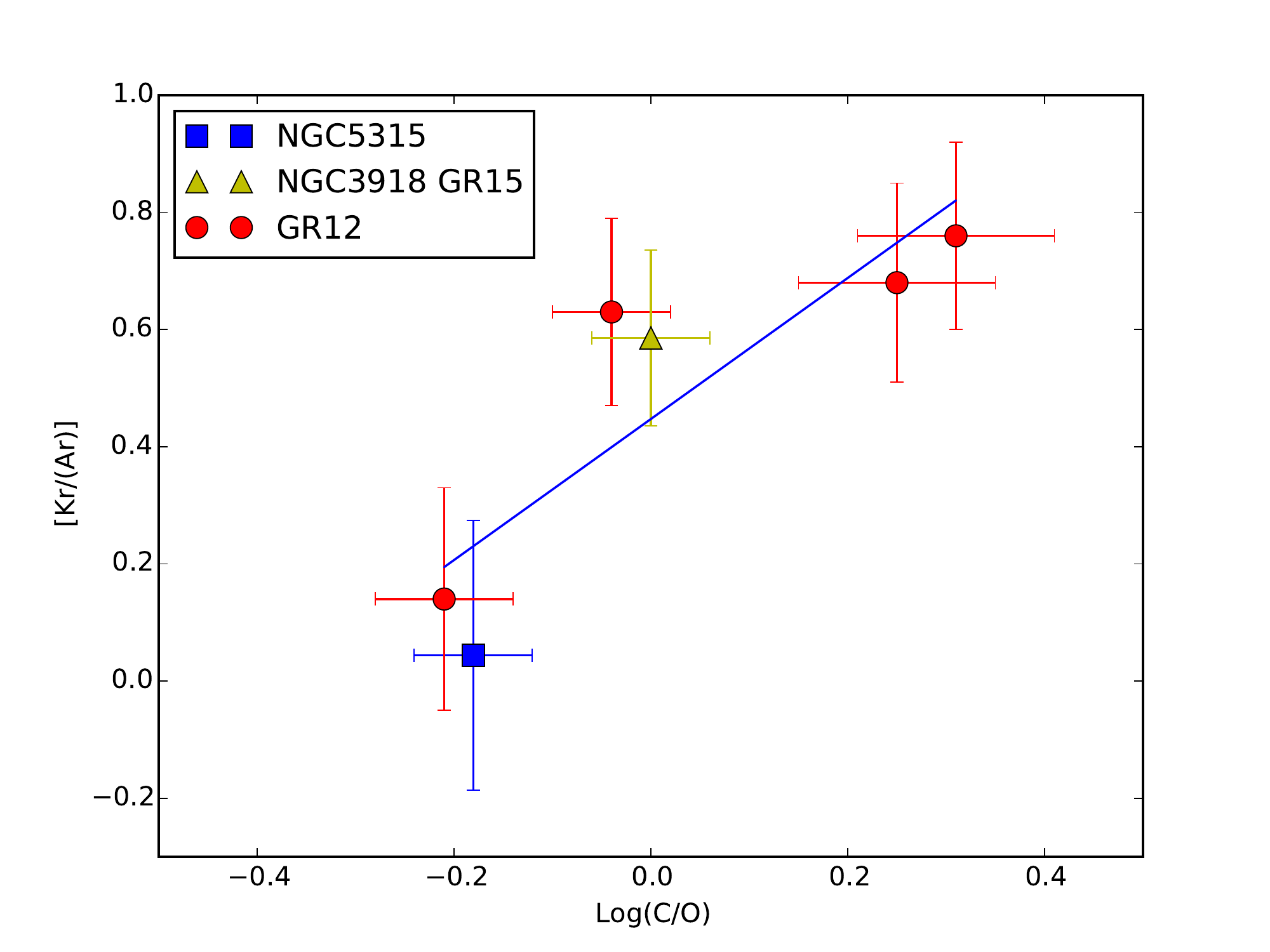}  
\caption{\label{kr_c} Correlation between Kr and C enrichment. Blue square is NGC 5315, yellow triangle is NGC 3918 (\cite{garciarojasetal15}) and red dots are the sample of \cite{garciarojasetal12}.).
}
\end{figure}

Finally we tested the theoretical conclusion that \emph{n}-capture element and C enrichment are strongly correlated, as they are processed in the same stellar layers and dredged up together to the surface during Third Dredge Up episodes (\cite{bussoetal01, karakaslattanzio14}). In Figure~\ref{kr_c} we plot our result for NGC\,5315 (blue square) along with the sample shown in figure 10 of \cite{garciarojasetal15}, finding a clear correlation. The C/O ratio is calculated using abundances from recombination lines. This result is very encouraging, but a bigger sample is strongly needed to further strengthen this predictions.

%
%
\small  
%
\section*{Acknowledgments}   
%
This work is based on observations collected at the European Southern Observatory, Chile, proposal number ESO 092.D-0189(A). This project is partially funded by the Spanish Ministry of Economy and Competitiveness under grant AYA2015-65205-P. JGR
acknowledges support from Severo Ochoa excellence program (SEV-2011-0187) postdoctoral fellowship. NCS gratefully acknowledges support of  this work from an NSF Astronomy and Astrophysics Postdoctoral Fellowship under award \hbox{AST-0901432.}

%

%
\end{document}